\documentclass[letterpaper, 10 pt, conference]{ieeeconf}
\IEEEoverridecommandlockouts
\usepackage{cite}
\usepackage{amsmath,amssymb,amsfonts,mathtools}
\usepackage{graphicx}
\usepackage{booktabs}
\usepackage{url}
\usepackage[hidelinks]{hyperref}

\newcommand{\Q}{\mathbb{Q}}

\newcommand{\T}{\mathbb{T}}
\newcommand{\U}{\mathbb{U}}
\newcommand{\sol}{\mathcal{R}}
\newcommand{\soli}{\mathcal{I}}
\newcommand{\code}[1]{\texttt{#1}}

\begin{document}

\title{\LARGE \bf Floating-point, exact and certified computation for control:\\
paradigms, pitfalls, and how to combine them}

\author{Christina Katsamaki and Fabrice Rouillier%
\thanks{C.~Katsamaki and F.~Rouillier are with Inria Paris, IMJ-PRG, Sorbonne Universit\'e,
Universit\'e Paris Cit\'e, CNRS, F-75005 Paris, France, and with PACE, a joint laboratory
of Inria and Safran.
{\tt\small \{christina.katsamaki, fabrice.rouillier\}@inria.fr}}}

\maketitle
\thispagestyle{empty}
\pagestyle{empty}

\begin{abstract}
Many questions in linear systems theory have a discrete answer: is a
polynomial Hurwitz, is a 2-D system structurally stable, is the peak gain
finite, which solution of a Riccati equation is stabilizing? Algorithms
implemented in floating point usually answer such questions, but often with a
limited, or even no, guarantee on the result, which depends on a tolerance
fixed a priori. Computer algebra and interval arithmetic provide other ways of
computing, which give guarantees on the result, but usually at a higher cost. In this paper, we
propose a short guide to these approaches: floating-point, exact,
multiprecision and interval arithmetic, certified root isolation and certified
Newton methods. Using small examples from control theory (each of them runs in
a few seconds), we show what each approach guarantees and in which situations
it fails, and how several approaches can be combined, for example by
certifying a numerical solution, or by reducing a problem exactly before
solving it with certified real solving. We then revisit the structural
stability of 2-D systems (with and without parameters) and the computation of
the $H_\infty$ norm, and we list the available implementations. The main
examples are given as reproducible worksheets based on the PACE ecosystem.
\end{abstract}

\section{Introduction}

Deciding whether a polynomial is Hurwitz, whether a transfer matrix has a
pole on the imaginary axis, whether a 2-D filter is structurally stable, or
which solution of an algebraic Riccati equation is the stabilizing one, comes
down to the sign of some quantity or to the existence of a real solution of
some system of equations. When this quantity vanishes, or when the solution
lies on a boundary, the usual numerical answer (an approximation compared
with a tolerance) becomes fragile. Of course, this answer is right in most
situations and this is the reason why it is used every day, but near limit
cases or singularities it may differ significantly from the correct result
without the user being warned.
In a recent campaign on $272$ systems \cite{HinfPeak}, we compared two
independent numerical routines for the $L_\infty$-norm. Both of them failed on
the question ``is there a pole on the imaginary axis'', but on opposite sides
(a large finite number where the norm is infinite, infinity where it is
finite) and never on the same instance, so that no choice of tolerance fixes both.

Computer algebra and interval arithmetic provide other ways of computing:
\begin{itemize}
\item exact arithmetic over the rational numbers decides a sign without any
  tolerance;
\item interval arithmetic computes enclosures which are guaranteed to contain
  the true value;
\item certified root isolation places each real root in a box containing
  no other root;
\item certified Newton methods prove that a solution exists near a
  floating-point approximation.
\end{itemize}
These tools are well known in computer algebra but, up to our knowledge, they
are rarely presented together with their limits and compared on the same
problems.

In this paper, we propose a short guide to these approaches, written for
people who use numerical tools for control and want to know what the other
tools can bring, when they fail and how to combine them. We proceed by
examples. Section~\ref{sec:paradigms} answers one question (is a given
polynomial Hurwitz?) with four kinds of arithmetic, Section~\ref{sec:pitfalls}
collects the pitfalls we met and Section~\ref{sec:combine} is devoted to
combinations of methods, on the algebraic Riccati equation. Section~\ref{sec:problems}
revisits the structural stability of 2-D systems, with and without parameters,
and the $H_\infty$ norm. The last section lists the implementations used and
gives an idea of the sizes of problems that can be handled.

The examples come from our own work on these questions
\cite{BQR19,BR16,BQRY21,QRY24,LRR27}. The implementations are those of the
PACE ecosystem\footnote{\url{https://pace.gitlabpages.inria.fr}}, and the main examples
are given as worksheets, in Julia and in Maple (Section~\ref{sec:implementations}).

Throughout the paper, the data are rational numbers (a decimal coefficient
such as $0.1$ is read as $1/10$). By \emph{certified}, we mean that the result
comes with a proof of its correctness (a sign that is decided, a box that
contains exactly one solution, a ball in which a solution exists).

\section{One question, four ways of computing}
\label{sec:paradigms}

A recurring question is whether a polynomial with exact coefficients is
Hurwitz, i.e.\ whether all its roots lie in the open left half-plane. Let us
take an example built to be hard for a floating-point method, with five times the same pair
of roots close to the imaginary axis:
\begin{equation}
  q(s) = \bigl(s^2 + a\,s + 1\bigr)^5, \qquad a = 2^{-10}.
  \label{eq:q}
\end{equation}
Every root has real part $-a/2 = -2^{-11}$, so the answer is yes. Note that all
the coefficients of $q$ are exactly representable in double precision: no
error comes from reading the data.

\subsection{Floating-point arithmetic}

The most common way to decide whether a polynomial is Hurwitz is to compute the eigenvalues of the companion matrix of
$q$ (this is what \code{roots} does in Matlab). In double precision, the
largest real part found is about $7\cdot 10^{-5}$. It is positive, and the
computed answer is thus ``unstable''. A
relative perturbation $\varepsilon$ of the coefficients moves a root of
multiplicity $m$ by about $\varepsilon^{1/m}$, here $(2^{-53})^{1/5} \approx
6\cdot 10^{-4}$, which is much more than the distance $2^{-11}$ between the
roots and the axis. This is not an error of the algorithm: it only ensures that
the computed roots are the exact roots of a polynomial whose coefficients are
very close to those of $q$ (this is called backward stability). Since $q$ is
very close to the stability limit, this nearby polynomial may have roots in the
right half-plane.

Changing the algorithm does not help in general. The Hurwitz criterion requires the
leading principal minors of the Hurwitz matrix of $q$ to be positive. Computed
in double precision from the same coefficients, the last two minors come out
negative, and the answer is again ``unstable''. A Routh table in double
precision happens to give the right answer on this example with pivots of
order $10^{-27}$, much smaller than the rounding errors: their sign proves
nothing. Without any other information, one cannot know which of the three
computations should be trusted.

\subsection{Exact arithmetic}

Over the rational numbers, the same Hurwitz criterion decides without any
tolerance. All the minors are positive, the last ones being of order
$10^{-43}$ and $10^{-70}$. The price to pay is the size of the numbers. Each
exact operation makes its data grow (a rational addition adds the sizes of its
operands, the entries of a determinant of order $k$ are about $k$ times longer
than the input), so the cost of an exact computation is the number of
operations multiplied by the size of the data each of them carries.
For the Hurwitz test, we use the Li\'enard--Chipart criterion by signed
subresultants. It decides
polynomials of degree $1000$ with coefficients of several thousand bits in
about one minute, where a rational elimination would not finish.

\subsection{Multiprecision and interval arithmetic}

Multiprecision floating-point arithmetic (the MPFR library \cite{FHLPZ07})
computes with as many bits as wanted, but increasing the length of the
mantissa does not tell whether the result is correct. This is what interval
arithmetic \cite{Moo66} provides: each number is replaced by an interval containing it, each operation rounds outwards, and a
sign is decided as soon as the interval does not contain $0$. With
multiprecision intervals (MPFI \cite{RR05}, or the ball arithmetic of Arb
\cite{Joh17}), one can then raise the precision until all the signs are
decided. On the minors of \eqref{eq:q}, $53$ bits decide the first four signs
and leave the other six undecided, four signs remain undecided at $106$ bits
and all of them are decided (positive) at $212$ bits. Interval arithmetic thus
never gives a wrong answer (it gives the right one or says that it does not
know), and its cost adapts to the difficulty of the input.

Interval arithmetic also has a limitation which has nothing to do with the
precision. When the same quantity enters a computation several times, the enclosure may
be much wider than the true range. It matters
as soon as the data are intervals themselves (Section~\ref{sec:pitfalls}).

\subsection{Certified methods}

The fourth way is to compute the result together with a proof. This can be done globally, by
locating all the roots, or locally, by certifying a given approximation.

\paragraph{Certified root isolation} Exact arithmetic first
removes the multiplicities: the square-free part $q/\gcd(q,q')$ has the same
roots, each of them simple, here $s^2 + 2^{-10}s + 1$. Its roots are then
isolated in disjoint boxes (for real roots) or discs (for
complex roots) that contain exactly one root each, and these boxes can be
refined to any precision. All our tools isolate real roots by Descartes' rule
of signs and bisection \cite{RZ04}, and complex roots by the method of
\cite{IM23}. Isolation assumes simple roots, so the square-free part must be
computed first, and exactly, since a multiple root of the exact polynomial
becomes a cluster of simple roots of the approximate one. On the example, the two discs lie in the open left
half-plane. Thus $q$ is Hurwitz, and we moreover know where its roots are.

For a system of polynomial equations with finitely many complex solutions, the
same scheme applies once the system has been reduced to a single variable.
This is what the Rational Univariate Representation (RUR) \cite{Rou99} does,
exactly: it provides a univariate polynomial $f(T)$ together with rational
functions $x_i = g_i(T)/g_0(T)$ such that the solutions of the system are the
images of the roots of $f$. The real solutions are the images of the real
roots of $f$: these are isolated as above, and the coordinates of the solutions
are enclosed in boxes
by evaluating the $g_i$ in interval arithmetic. For two equations in two
unknowns, the RUR comes from subresultants, without any Gr\"obner basis
\cite{BLPR15}; this is the solver of Section~\ref{sec:problems}.

\paragraph{Certified Newton methods}
These methods start from an approximate solution of a system
$F(x) = 0$ and prove that an exact solution lies nearby. Kantorovich's
theorem \cite{KA64} gives such a proof from the values of $F$, of its Jacobian
matrix and of a bound on its second derivatives at the point: if a constant
$C$ computed from these quantities is at most $1$, then there is exactly one
solution in a ball centred at the point, with a computable radius. Interval Newton methods
\cite{Moo66} go further and contract a box known to contain a solution, each
step that maps the box into its own interior being a proof of existence and
uniqueness. These methods are local (they say nothing about the other
solutions), but their cost is polynomial in the size of the system,
since they only evaluate the system and its derivatives at one point.

To summarize, on \eqref{eq:q}, double precision answers ``unstable''
(wrongly), exact arithmetic and certified isolation answer ``stable'' with a
proof, and interval arithmetic answers ``stable'' once the precision is
sufficient and ``don't know'' before. Basically, exact arithmetic and certified
isolation are global and always decide, interval arithmetic decides or
abstains, while Newton methods only certify a given solution. In the next
sections, we show how each of them can fail and how they can be combined.

\section{Pitfalls}
\label{sec:pitfalls}

\subsection{The data are not floating-point numbers}

Let $\delta = 2^{-20}$ and $p(s) = ((s+\delta)^2+1)^4$. This polynomial is
Hurwitz, its roots having real part $-\delta$, and its coefficients are
rational numbers whose denominators reach $161$ bits. Reading them in
double precision changes the polynomial, and the exact test shows that the
approximate polynomial is not Hurwitz. This is the reason why the exact and certified tools take rational
numbers as input.

\subsection{A tolerance near the boundary}

Consider the 2-D polynomial $D_\varepsilon = 2 + z_1 + z_2 + \varepsilon$.
When $\varepsilon > 0$, the system is structurally stable (see
Section~\ref{sec:2d} for the definition) since $|z_1 + z_2| \le 2$ on the
closed unit bidisc. When $\varepsilon < 0$, it is not, since $D_\varepsilon$
vanishes at $z_1 = z_2 = -1 - \varepsilon/2$. Take $\varepsilon = \pm
10^{-12}$. A floating-point test would sample $|D_\varepsilon|$ on the torus
$|z_1| = |z_2| = 1$, and on a grid of $1000 \times 1000$ points the minimum is
$10^{-12}$ in both cases, so that the stable case cannot be distinguished
numerically from the unstable one. The exact reduction of Section~\ref{sec:2d} turns the
question into the existence of a real solution of a bivariate system, and
after elimination the univariate polynomial has its roots at distance about
$\sqrt{|\varepsilon|}/2$ (i.e.\ $5\cdot 10^{-7}$) from the real axis. A
floating-point root finder, which declares real every root whose imaginary part
is below a tolerance, answers ``two real roots'' for both signs of $\varepsilon$ when the tolerance
is $10^{-6}$, and gives the right answer when it is $10^{-8}$. The exact test
finds no zero on the torus for $\varepsilon > 0$ and two of them for
$\varepsilon < 0$, without any tolerance.

\subsection{Sampling the frequencies}

The $H_\infty$ norm of a stable system is its largest amplification over all
frequencies (Section~\ref{sec:hinf}). For a lightly damped oscillator
$G(s) = 1/(s^2 + 2\zeta s + 1)$ with $\zeta = 10^{-3}$, the peak is narrow and
its value is known in closed form (about $500$). Sampling $|G(i\omega)|$ at
$200$ frequencies logarithmically spaced between $10^{-2}$ and $10^{2}$ gives $22$, simply
because the resonance is narrower than the step of the grid, and the result
does not show it. With $\zeta = 0$, the norm is infinite
and the same sampling returns a finite number. In the first case, the
certified methods of Section~\ref{sec:hinf} enclose the norm between two
rationals; in the second one, they decide that a pole lies on the axis.

\subsection{Intervals and uncertain data}

When the coefficients come from measurements, they are only known to lie in
intervals, which corresponds to an infinite family of polynomials. The
question becomes whether every polynomial of the family is Hurwitz. A straightforward approach would be to compute the Hurwitz minors
with interval coefficients, but this fails quickly. On $(s+1)^6$ with each
coefficient varying independently by $\pm 12.5\%$, three signs remain
undecided, and raising the precision from $53$ to $1000$ bits changes nothing
since the width of the intervals now comes from the data and is overestimated
by interval arithmetic. Kharitonov's theorem \cite{Kha78} gives another way to
answer. Provided that
the interval of the leading coefficient does not contain $0$, every polynomial
of the family is Hurwitz if and only if four particular polynomials, built
from the bounds of the intervals, are. With rational bounds, these four tests
are exact. Here they prove that the family is stable, and with $\pm 25\%$ two
of the four polynomials are not Hurwitz, so that the family is not stable. In
both cases, the interval computation gave no wrong answer and only failed to
decide.

\subsection{A certificate depends on the scaling of the equations}

The constant $C$ of Kantorovich's theorem depends on the way the equations are
written, so a certificate may fail at a point which is a perfectly good
approximation. On the Riccati equations of Section~\ref{sec:combine}, the test
at the floating-point solution fails from dimension $n = 12$ on ($C = 16$).
Multiplying the system by an approximate inverse of its Jacobian matrix at the
point does not change the solutions but keeps the cancellations between the
second derivatives, and the same points get certified up to $n = 13$, for
about three times the cost of the test.

\section{Combining the approaches}
\label{sec:combine}

The approaches are not exclusive and, in practice, the most useful
computations combine them. We show this on the continuous algebraic Riccati equation (CARE)
\begin{equation}
  A^T X + X A - X B B^T X + Q = 0,
  \label{eq:care}
\end{equation}
which is usually solved by a double-precision floating-point algorithm (\code{care} in Matlab, or
Kleinman's iteration \cite{Kle68}). Our example is the system $\dot x = Ax +
Bu$ with one input,
$A = \begin{psmallmatrix} 0 & 1\\ 2 & -1\end{psmallmatrix}$,
$B = \begin{psmallmatrix} 0\\ 1\end{psmallmatrix}$, $Q = I$. Without control
it is unstable. The control minimizing $\int_0^\infty (x^Tx + u^2)\,dt$ is
$u = -B^TXx$, where $X$ is the solution of \eqref{eq:care} that makes
$A - BB^TX$ stable (the stabilizing solution).

\subsection{Certify, then refine, a numerical solution}

Started from a stabilizing gain, Kleinman's iteration converges quadratically
to the stabilizing solution \cite{Kle68}. In double precision, it gives a
matrix $X_f$ whose residual in \eqref{eq:care} is of order $10^{-15}$. Such a
small residual only says that $X_f$ solves a nearby equation.

Writing $X = \begin{psmallmatrix} x_1 & x_2\\ x_2 & x_3\end{psmallmatrix}$,
\eqref{eq:care} becomes a system of three quadratic equations in three
unknowns with rational coefficients. We run the Kantorovich test at $X_f$ in
interval arithmetic and get $C \approx 5\cdot 10^{-14}$, far below $1$: an
exact solution lies within $4\cdot 10^{-15}$ of $X_f$. The test costs one
evaluation of the system, of its Jacobian matrix and of a bound on its second
derivatives, at a single point. Interval Newton, started from the box given by the Kantorovich ball (from
the point, it would only compute a Newton step), then refines the solution
to $200$ bits. On this example the exact solution is known ($x_1 = 7 + \sqrt5$, $x_2 =
2 + \sqrt5$, $x_3 = \sqrt5$) and it does lie in the boxes.

\subsection{Solve exactly, to know that nothing else exists}

The certificate is local. In order to know that the certified solution is the
only stabilizing one, we compute all the real solutions of the system by the
RUR and certified isolation, and we find four of them. The closed loop
$A - BB^TX$ has characteristic polynomial $s^2 + (1 + x_3)s + (x_2 - 2)$,
whose coefficients are known as intervals at each solution, and the Hurwitz
test on interval coefficients decides that exactly one solution is
stabilizing. Its box lies in the Kantorovich ball around $X_f$.

\begin{table}[t]
\caption{CARE of dimension $n$: Kantorovich constant at the Kleinman
solution without ($C$) and with ($C_p$) preconditioning, time of Kleinman
plus the test ($t_c$), real solutions and time of exact solving ($t_e$), in
seconds}
\label{tab:care}
\centering
\footnotesize
\setlength{\tabcolsep}{4pt}
\begin{tabular}{@{}rrllrrr@{}}
\toprule
$n$ & unknowns & $C$ & $C_p$ & $t_c$ & real sol. & $t_e$ \\
\midrule
 2 &   3 & $3\cdot 10^{-14}$ & $9\cdot 10^{-15}$ & 0.0002 &   4 & 0.0005 \\
 6 &  21 & $3\cdot 10^{-9}$  & $1\cdot 10^{-10}$ & 0.02   &  16 & 0.01 \\
 8 &  36 & $8\cdot 10^{-6}$  & $2\cdot 10^{-7}$  & 0.12   &  32 & 0.18 \\
10 &  55 & $9\cdot 10^{-4}$  & $2\cdot 10^{-5}$  & 0.43   &  64 & 5.2 \\
12 &  78 & $16$              & $0.011$           & 1.3    & 128 & 973 \\
13 &  91 & $5\cdot 10^{4}$   & $0.13$            & 2.2    & --  & -- \\
14 & 105 & $6\cdot 10^{14}$  & $5.6$             & 3.6    & --  & -- \\
\bottomrule
\end{tabular}
\end{table}

In order to see what happens when the size grows, Table~\ref{tab:care}
considers a family of such systems of dimension $n$ built in the same way (one input, $Q = I$) and
gives the Kantorovich constant at the Kleinman solution, the cost of the
certificate, the number of real solutions and the cost of exact solving. Up to $n \approx 8$, solving the whole system exactly is as cheap as
certifying a single solution: the test bounds second derivatives over
$n(n+1)/2$ unknowns, and its cost grows quickly too. Beyond, exact solving
becomes the expensive one. Its cost follows the number of complex solutions
counted with multiplicities, which doubles every two dimensions, while the certificate remains affordable
(a few seconds at $n = 14$), the preconditioned test keeping it valid beyond
$n = 11$. In practice, this suggests using the certificate in order to trust a
numerical result and, when the size allows it, exact solving in order to know
that there is no other solution.

\subsection{Uncertain data}

Suppose now that the entries of $A$ are only known up to $\pm\delta$, and that
we want to know whether the stabilizing solution resists the perturbations.
Solving the family of systems thus represented means computing boxes,
common to all its instances, each of which contains exactly one real solution
of each instance. In practice, we solve
exactly the system at the nominal data, then run the Kantorovich test with
interval coefficients around each solution, so that the constants are valid
for every system of the family. On the $2 \times 2$ example, one test per
solution certifies the four common boxes up to $\delta \approx 8\cdot 10^{-4}$
(about $2\cdot 10^{-3}$ with preconditioning). For $\pm 1\%$, the box of
parameters is cut into $625$ sub-boxes, the tests run in about one second and
the stabilizing solution is kept everywhere. The method fails when a
solution becomes singular for some instance.

\begin{figure}[t]
\centering
\includegraphics[width=\columnwidth]{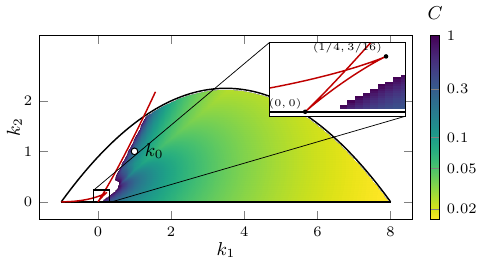}
\caption{Gains of a PI controller certified stable from $k_0$, coloured by the
Kantorovich constant $C$; black: boundary of the exact region; red: double
roots of the closed loop, with two cusps (inset)}
\label{fig:paving}
\end{figure}

The same test applies to a whole box of parameters when they enter the
coefficients, so that one can pave the stable gains around a known stable
design. Take the plant $1/(s+1)^3$ with a PI controller $k_1 + k_2/s$, i.e.\
$p(s) = s^4 + 3s^3 + 3s^2 + (1+k_1)s + k_2$. Starting from the box of a stable
gain $k_0$, each box is accepted when the Kantorovich tests on the four roots of $p$,
with $1+k_1$ and $k_2$ as intervals, give disjoint balls in the open left
half-plane, and its neighbours are then tested. The computed boxes cover $88\%$ of the exact region (Figure~\ref{fig:paving}). Near the boundary of the
region, $C$ stays small and the stability test stops the paving, while $C$
tends to $1$ near the red curve, where $p$ has a double root: the systems
are still stable there, but the root is singular and Newton fails.

\section{Two problems revisited}
\label{sec:problems}

\subsection{Structural stability of 2-D systems}
\label{sec:2d}

A discrete 2-D system with transfer function $N/D$, $D \in \Q[z_1,z_2]$, is
structurally stable when $D$ has no zero on the closed unit bidisc
$\overline{\U}^2 = \{|z_1| \le 1, |z_2| \le 1\}$ \cite{BQR19}. Since this is a
question about infinitely many points, numerical strategies rely on sampling.
According to \cite{DMS77}, $D$ is stable if and only if the univariate
polynomials $D(z_1, 1)$ and $D(1, z_2)$ have no root in the closed unit disc
and $D$ has no zero on the torus $\T^2 = \{|z_1| = |z_2| = 1\}$. In
\cite{BQR19,BR16}, we reformulated these conditions so that certified tools can
be used.

The two univariate conditions become Hurwitz tests after the change of
variable $z = (1-s)/(1+s)$, which maps the open left half-plane onto the
exterior of the unit disc. For the torus, we use $z = (x - i)/(x + i)$, which maps
the real line onto $\T \setminus \{1\}$. Substituting it for $z_1$ and $z_2$
and clearing the denominators, the numerator of $D$ reads $\sol(x_1,x_2) +
i\,\soli(x_1,x_2)$ with $\sol, \soli \in \Q[x_1,x_2]$, and $D$ has no zero on
the torus if and only if the bivariate system $\sol = \soli = 0$ has no real
solution (the points with a coordinate equal to $1$ are covered by the
univariate tests). This system is solved by computing a bivariate RUR,
followed by certified isolation of the real roots.

As an example, take $D = (2z_1^2 + 10z_1 + 12) + (z_1^2 + 5z_1 + 6)z_2$, i.e.\
$(z_1+2)(z_1+3)(z_2+2)$, a form that the algorithms do not use. The two
univariate tests pass. We get $\sol = 36x_1^2x_2 - 10x_1 - 6x_2$ and $\soli =
12x_1^2 + 30x_1x_2 - 2$. The univariate polynomial of the RUR is $324T^4 +
369T^2 + 100 = (36T^2 + 25)(9T^2 + 4)$, and it has no real root. Thus $D$ is
structurally stable and each step of the computation is exact or certified. On the near-boundary
example of Section~\ref{sec:pitfalls}, the two zeros found for $\varepsilon <
0$ are isolated by boxes of width $10^{-30}$, at distance about $10^{-6}$ from
$z_1 = z_2 = -1$.

The cost is that of the bivariate solver, which depends on the degree of $D$:
for a dense $D$ of degree $40$
in each variable, the RUR takes about two seconds, and the isolation from one
second, when there is no real solution, to about a minute and a half when
there are several hundreds of them.

\subsection{With parameters}
\label{sec:param}

When $D$ depends on parameters $u = (u_1, u_2)$, the set of $u$ for which
the system is stable is a union of regions of the parameter plane bounded by
algebraic curves. We compute these curves exactly, then decide stability at
one rational point of each region with the test above \cite{BR16}. The number of real solutions of $\sol = \soli = 0$ is locally
constant outside the discriminant variety of the system with respect to $u$
\cite{LR07}, and the univariate Hurwitz conditions can only change sign on
the zero sets of the polynomials in $u$ given by the parametric
Li\'enard--Chipart criterion. The stability boundary is thus contained in the union of
these curves.

One point per region is obtained either by a partial cylindrical algebraic
decomposition \cite{Col75} adapted to the curves, which gives one rational
point in each cell of maximal dimension but cuts the regions into more cells
than needed, or from the certified topology of the curve \cite{Bou14}, which
gives the connected components of its complement with a rational sample point
in each; we use the latter to draw the regions.

\begin{figure}[t]
\centering
\includegraphics[width=0.48\columnwidth]{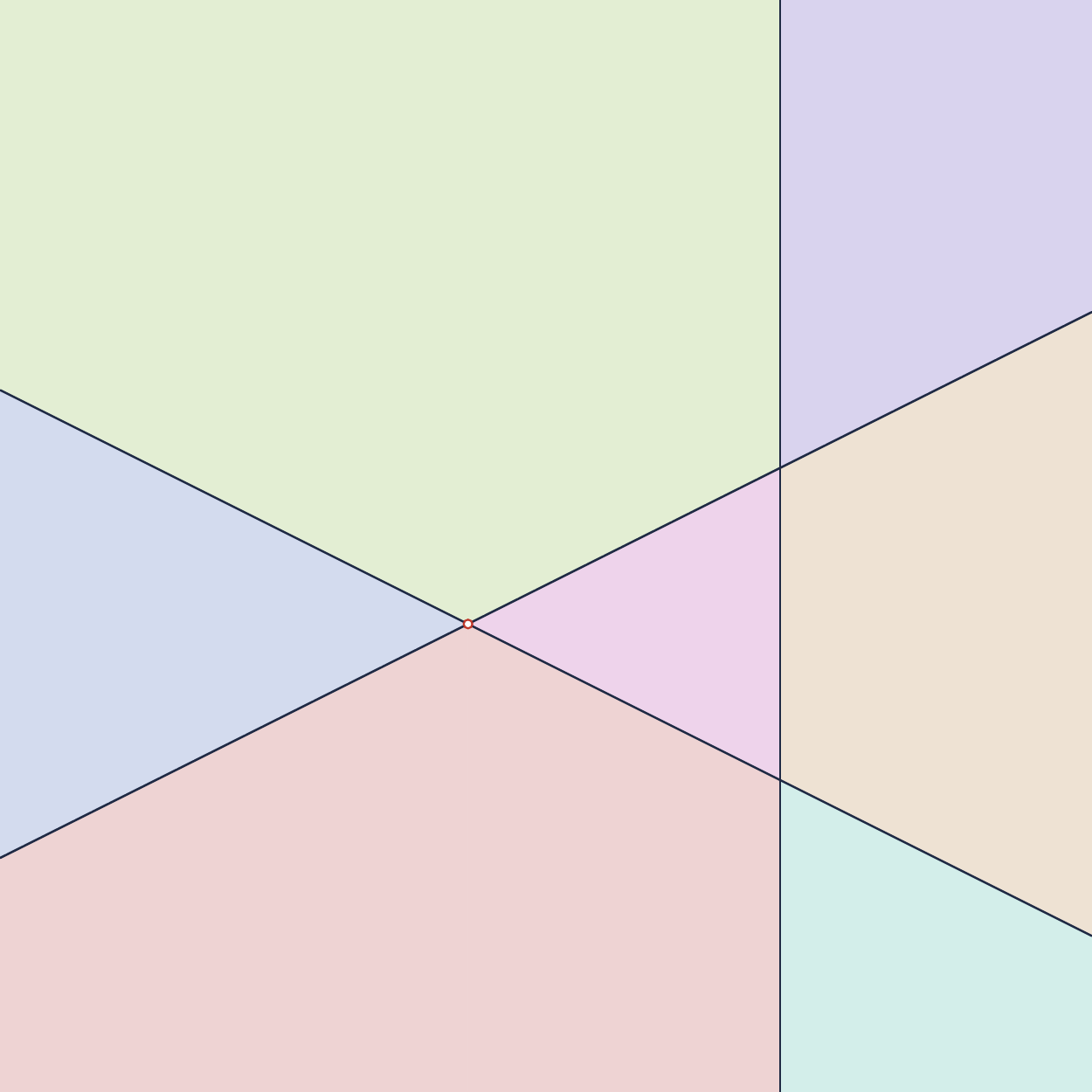}\hfill
\includegraphics[width=0.48\columnwidth]{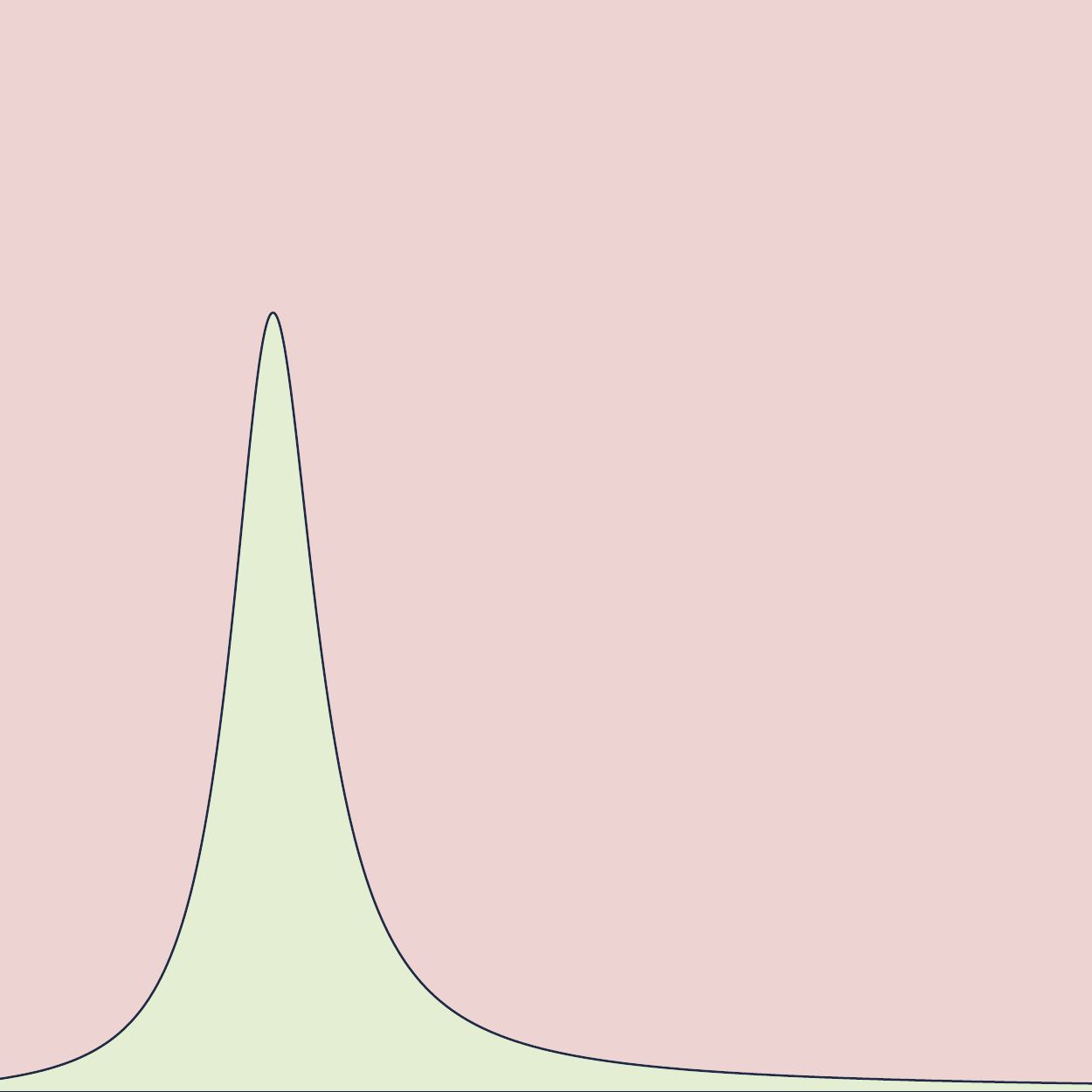}
\caption{Left: the seven regions of the $(u_1,u_2)$ plane for the example of
\cite{LH21}; the stable one is the triangle containing $(0,-1)$. Right: the
curve $c(\Gamma, \Omega) = 0$ of the COMPleib model AC1, $\Omega = \omega^2$
horizontal, $\Gamma$ vertical; the $H_\infty$ norm is the square root of the
highest point of the branch}
\label{fig:regions}
\end{figure}

On the example $D = u_2z_1z_2 - u_1z_2 - u_2z_1 + z_1z_2 - z_1 + 1$ of
\cite{LH21}, the stability boundary consists of three lines. The decomposition has $12$
cells and the complement of the curve has $7$ regions
(Figure~\ref{fig:regions}, left); in both cases, exactly one of them is
stable, namely the open triangle $u_1 < 1$, $u_1 - 2u_2 > 1$, $u_1 + 2u_2 >
-3$. A grid of the parameter plane would only give a picture, which may miss the
regions smaller than its step, while we get a finite list of regions, each of them
certified stable or unstable. When the boundary is a curve of
total degree about $100$, the regions are obtained in a few seconds; for a
total degree of $250$, it takes about one minute.

\subsection{The $H_\infty$ norm}
\label{sec:hinf}

For a stable system with transfer matrix $G(s)$, the $H_\infty$ norm is the
peak gain $\gamma = \sup_{\omega} \overline\sigma(G(i\omega))$, where
$\overline\sigma$ denotes the largest singular value. Numerically, it is
computed by the bisection of \cite{BBK89} and its variants (\code{hinfnorm}
and \code{getPeakGain} in Matlab), all of which need a tolerance. Following
\cite{KS06}, we showed in \cite{BQRY21} that $\gamma^2$ is the largest
critical value, on the branch $\Omega \ge 0$, of the plane algebraic curve
$c(\Gamma, \Omega) = 0$ which is the numerator of $\det(\Gamma I -
G^T(-s)G(s))$ with $s^2 = -\Omega$ (or a limit at $\Omega = 0$ or $\infty$);
the parametric case is treated in \cite{QRY24}. The critical points are the
real solutions of a bivariate system, which the RUR and certified isolation
enclose. Hence $\gamma^2$ is isolated between two rationals, and whether a
pole lies on the imaginary axis is decided by a gcd and a real root count. Two
implementations are available, by Sturm--Habicht sequences on the transfer
matrix \cite{BQRY21} or by building the curve directly from a state-space
realization \cite{HinfPeak}; we use the second one below.

On the oscillator of Section~\ref{sec:pitfalls}, both algorithms isolate the closed-form
value $\gamma \approx 500.00025$ (the sampling was $20$ times too small), and
with $\zeta = 0$ they decide that the norm is infinite. On the aircraft model
AC1 of COMPleib \cite{Lei04} ($5$ states, $3$ inputs, $2$ outputs), the
sampling is $0.6\%$ below the certified value $\gamma \approx 2.1693$,
attained at $\omega^2 \approx 0.0334$ (this error is only known thanks to the
certified value). The curve behind this norm has
degrees $2$ and $4$ in $\Gamma$ and $\Omega$, with integer coefficients of up
to $93$ digits, and Figure~\ref{fig:regions} (right) is its certified
drawing.

Concerning the cost, the COMPleib models with up to ten states (about seventy
of them) are all treated in a few seconds, and the certified norm agrees with
the numerical routines except where the latter decide wrongly on the axis
\cite{HinfPeak}. On single-input single-output families, the norm is
certified in about two minutes at order $1000$, and in about a minute and a half for $4 \times 4$ transfer matrices
of order $150$.

\section{Implementations and orders of magnitude}
\label{sec:implementations}

\begin{table}[t]
\caption{Implementations used in this paper}
\label{tab:impl}
\centering
\footnotesize
\setlength{\tabcolsep}{4pt}
\begin{tabular}{@{}lp{5.3cm}@{}}
\toprule
Tool & What it provides \\
\midrule
\multicolumn{2}{@{}l}{\emph{PACE (Julia packages and a Maple toolbox on the same C library)}} \\
LACE & RUR, real root isolation, bivariate solver, Kantorovich test,
       interval Newton, discriminant variety ideals \\
MPFI.jl & multiprecision interval arithmetic \cite{RR05} \\
PWPoly.jl & certified complex roots \cite{IM23} \\
DiscriminantVariety.jl & discriminant varieties \cite{LR07} \\
CurveTopology.jl & topology, drawing and complement of plane curves \\
PACE.jl & partial CAD, Hurwitz tests, utilities \\
PaceControl.jl & 2-D stability with parameters, $H_\infty$ norm
                 (\code{Hinf\_StHa}, \code{hinf\_peak}) \\
\bottomrule
\end{tabular}
\end{table}

Table~\ref{tab:impl} lists the implementations. All the computations of this
paper run through the PACE packages, which share a C library (LACE) built on
GMP, MPFR and MPFI for the multiprecision and interval parts and on
multimodular computations for the exact parts. The same library is available
as a Maple toolbox, and the five worksheets (the Hurwitz test four ways, the
Riccati equation, 2-D stability, the $H_\infty$ norm, the stability regions)
exist in both languages, with the same conclusions; each of them runs in less
than a minute on a laptop. The uncertain data of Section~\ref{sec:combine},
the paving of Figure~\ref{fig:paving} and the orders of magnitude below come
from separate Julia scripts.

\begin{table}[t]
\caption{Orders of magnitude on a laptop (one thread unless stated)}
\label{tab:sizes}
\centering
\footnotesize
\setlength{\tabcolsep}{4pt}
\begin{tabular}{@{}lll@{}}
\toprule
Problem & Size & Time \\
\midrule
Exact Hurwitz test & degree $200$ / $1000$ & $0.2$ s / $70$ s \\
Real root isolation & degree $500$, $2600$ bits & $110$ s \\
2-D stability, bivariate RUR & degree $40$ ($3200$ sol.) & $2$ s \\
\quad then isolation & $0$ / $450$ real sol. & $1.5$ s / $90$ s \\
Stability regions, topology & total degree $100$ / $250$ & $4$ s / $45$ s \\
CARE, certificate & $n = 10$ / $14$ & $0.4$ s / $4$ s \\
CARE, all real solutions & $n = 10$ / $12$ & $5$ s / $16$ min \\
$H_\infty$ norm, SISO & order $200$ / $1000$ & $1$ s / $2$ min \\
$H_\infty$ norm, $4\times4$ & order $50$ / $150$ & $2$ s / $80$ s \\
\bottomrule
\end{tabular}
\end{table}

Table~\ref{tab:sizes} gives orders of magnitude for the sizes that can be
handled. They come from a single run on a laptop, on the examples of this
paper and on random families of the same kind. The exact and certified
methods are not meant to replace the numerical ones at the sizes where the
latter are routine (state dimensions in the thousands for the $H_\infty$
norm, for instance). At such sizes, the exact computations do not finish and
one should certify the numerical solution.

\section{Conclusion}

We have presented four ways of computing on questions of control theory that
have a discrete answer, and the examples show that none of them is sufficient
on its own. In practice, we recommend keeping the numerical solver and adding
what is missing: a certificate when a result must be trusted, exact
enumeration when one must know that nothing else exists, and an exact
reduction of the problem (a change of variables, a square-free part, a
resultant) before the certified solving. The worksheets given with this paper
can be used as a starting point to try these combinations on other problems.

\bibliographystyle{IEEEtran}
\bibliography{refs}

\end{document}